\documentclass[preprint]{revtex4} 
\usepackage{epsfig}
\usepackage{graphicx}% Include figure files
\usepackage{dcolumn}% Align table columns on decimal point
\usepackage{bm}% bold math
\begin{document}
\title{Universal behavior of optimal paths in weighted networks\\with general
  disorder}

\author{Yiping Chen$^1$, Eduardo L\'{o}pez$^{1,2}$, Shlomo Havlin$^{1,3}$ and
  H.Eugene Stanley$^1$}

\affiliation{$^1$Center for Polymer Studies, Boston University, Boston,
  Massachusetts 02215, USA\\$^2$Theoretical Division, Los Alamos National
  Laboratory, Mail Stop B258, Los Alamos, NM 87545, USA\\$^3$Minerva Center
  of Department of Physics, Bar-Ilan University, Ramat Gan, Israel}


\begin{abstract}
  We study the statistics of the optimal path in both random and scale free
  networks, where weights $w$ are taken from a general distribution $P(w)$.
  We find that different types of disorder lead to the same universal
  behavior. Specifically, we find that a single parameter ($S \equiv
  AL^{-1/\nu}$ for $d$-dimensional lattices, and $S\equiv AN^{-1/3}$ for
  random networks) determines the distributions of the optimal path length,
  including both strong and weak disorder regimes. Here $\nu$ is the
  percolation connectivity exponent, and $A$ depends on the percolation
  threshold and $P(w)$.  For $P(w)$ uniform, Poisson or Gaussian the
  crossover from weak to strong does not occur, and only weak disorder
  exists.
\end{abstract}

\maketitle
%{\setlength{\baselineskip}{2\baselineskip}
%INTRODUCTION

The study of the optimal path in disordered networks has attracted much
interest in recent years, due to its relation to many complex systems of
interest, including polymers~\cite{cieplak,rintoul}, nanomagnet
transport~\cite{strelniker}, surface growth~\cite{barabasi-stanley}, spin
glasses~\cite{void}, tumoral growth immune response~\cite{bru}, and complex
networks~\cite{lidia}. The disorder is realized by assigning random weights
$w$ to the links of the network, where the weights are drawn from a
distribution $P(w)$, chosen to reflect the characteristics of the system
under study. Examples include exponential disorder for tunnelling
effects~\cite{strelniker}, power law disorder for driven diffusion in random
media~\cite{alex}, lognormal distribution in conductance of quantum
dots~\cite{edward} and Gaussian distribution for polymers~\cite{redner}.

Of particular current interest is the scaling behavior of the optimal path
length $\ell$ for different weight distributions~\cite{lidia,alex,cieplak}.
The optimal path between two nodes is the path that minimizes the sum of the
weights along the path. By definition, the \emph{weak disorder limit} is
obtained when almost all links contribute to the total weight of the optimal
path. For this case $\ell \sim L$~\cite{redner} for a $d$-dimensional lattice
of linear size $L$, and $\ell \sim \mathrm{log}N$ for random networks of $N$
nodes~\cite{lidia}. The \emph{strong disorder limit} is obtained when a
single link weight dominates the sum of weights along the
path\cite{footnote1}. For this case
\begin{equation}
\ell\sim L^{d_{opt}},
\end{equation}
with $d_{opt}=1.22 \quad (d=2)$ and $d_{opt}=1.42 \quad (d=3)$ for
lattices~\cite{cieplak,porto}, and $\ell \sim N^{1/3}$ for random
networks~\cite{lidia}. It is commonly agreed that strong disorder arises
only when $P(w)$ is broad enough.

Lacking is a general criterion to determine which form of $P(w)$ can lead to
strong disorder, and a general condition when strong or weak disorder occurs.
Here we derive analytically such a criterion, and support our derivation by
extensive simulations. Using this criterion we show that certain power law
distributions and lognormal distributions $P(w)$ can lead to strong disorder
and to a weak-strong disorder crossover~\cite{porto}.  We also show that for
$P(w)$ uniform, Poisson or Gaussian, only weak disorder occurs regardless of
the broadness of $P(w)$. Importantly, we find that for all $P(w)$ that
possess a strong-weak disorder crossover, the distributions $p(\ell)$ of the
optimal path lengths display the same universal behavior.

  % SIMULATION
A useful way to study properties of the $P(w)$ is to relate it to the uniform
distribution $\pi(x)=1$, with $x\in [0,1)$.  From the relation
$P(w)dw=\pi(x)dx=dx$, $w$ can be obtained through the ``disorder function''
$f(x)$ defined as $w\equiv f(x)$, which is equivalent to
\begin{equation}  
x \equiv f^{-1}(w) =\int_0^w P(w')dw'.
\label{fx}
\end{equation}
In this form, every link has assigned to it a number $x$ between 0 and 1, and
its corresponding weight is given by the relation $w=f(x)$.  Using $f(x)$, we
can write the total weight $w_{opt}\equiv \Sigma ^{\ell} _{j=1}w_j$ of the
optimal path as
\begin{equation}
  w_{opt}=f(x_1)+f(x_2)+\ldots +f(x_{\ell}),
\label{eopt}
\end{equation}
where the terms of the sum are arranged in decreasing order of values of $w_j
\equiv f(x_j)$.

Next, we find the condition for strong disorder. We consider the first two
terms of Eq.~(\ref{eopt}), since for strong disorder only the first term of
~(\ref{eopt}) is dominant~\cite{perlsman}. The reason for choosing only two
terms is that if the second term is very small compared to first then all
others can be neglected and we are in the strong disorder, while if the
second term cannot be neglected all others also cannot be neglected and we
are in the weak disorder. Thus only the first two terms determine the
disorder. By definition $w_2/w_1=f(x_2)/f(x_1)<1$, so we can expand $f(x_2)$
around $x_1$,
\begin{equation}
  w_1+w_2= f(x_1)\left[1+\frac{f(x_2)}{f(x_1)}\right]\simeq f(x_1)[2-S],
\label{eopt2}
\end{equation}
where
\begin{equation}
S\equiv \left.\frac{d(\mathrm{ln}f)}{dx}\right\vert_{x=x_1}(x_1-x_2).
\label{tau}
\end{equation}
To see that $S$ determines the disorder strength and the weak-strong disorder
crossover, we note that $(i)$ for the \emph{strong disorder limit},
$w_{opt}\approx f(x_1)$, and therefore $f(x_2)/f(x_1)\rightarrow 0$. Since
$S\simeq 1-f(x_2)/f(x_1)$, it follows that $S\rightarrow 1$, and $(ii)$ for
the \emph{weak disorder limit} all weights $w_j$ contribute to $w_{opt}$, so
$f(x_1)\gtrsim f(x_2)$, and $S\ll 1$.

To identify $x_1$ and $x_2$ in ~(\ref{tau}), we show that the strong disorder
case is related to percolation~\cite{porto}. The percolation problem
corresponds to the construction of random networks by the removal of links
with probability $1-p$. The removal of links relates to the strong disorder
limit, because all links with weights $w_j$ larger than necessary to keep the
network connected are not relevant and can be removed~\cite{footnote}.
Therefore, the weights that are typically removed are those satisfying
$w>f(p_c)$, where $p_c$ is the critical percolation threshold, the value of
$p$ for which the system switches between being connected to disconnected.
Hence the value of $x_1$ should be $p_c$ (in the limit of infinite system
size).

For percolation on a \emph{finite} system, $p_c$ is not unique but follows a
distribution characterized by a width which scales as $L^{-1/\nu}$, where
$\nu$ is the percolation connectivity length exponent~\cite{stauffer,bunde}.
Thus for strong disorder, both $x_1$ and $x_2$ are distributed close to $p_c$
with widths $L^{-1/\nu}$. For typical values of $x_1$ and $x_2$, $x^*_1$ and
$x^*_2$ respectively, we expect
\begin{equation}
\frac{x^*_1 - x^*_2}{p_c} \sim L^{-1/\nu}.
\label{L}
\end{equation}
Combining Eqs.~(\ref{tau}) and (\ref{L}), we find
\begin{equation}
S \sim p_c\frac{ d(\mathrm{ln}f)}{dx}\mid_{x=p_c}L^{-1/\nu}.
\label{tau2}
\end{equation}
Using $w=f(x)$, and $f'(x)=1/P(w)\vert_{w=f(x)}$, Eq.~(\ref{tau2}) can be
expressed as
\begin{equation}
S\sim \frac{p_c L^{-1/\nu}}{w_c P(w_c)}\equiv AL^{-1/\nu},
\label{tau3}
\end{equation}
where $w_c$ is the solution of the equation
\begin{equation}
\int_0^{w_c}P(w)dw=p_c.
\label{tau4}
\end{equation}
Equation ~(\ref{tau3}) expresses $S$ as a function of
the general disorder distribution $P(w)$ and $L$. 

Next we extend our theory for $S$ to random networks. Percolation at
criticality on Erd\H{o}s-R\'{e}nyi (ER) networks~\cite{erdos} is equivalent
to percolation on a lattice at the upper critical dimension
$d_c=6$~\cite{lidia,cohen}. For $d=6$, $L \sim N^{1/6}$, and $\nu=1/2$. Thus,
using (\ref{tau3}) we obtain,
\begin{equation}
S\sim AL^{-1/\nu}=AN^{-1/6\nu}=AN^{-1/3}.
\label{er}
\end{equation}
Thus, for ER networks we simply replace $L^{-1/\nu}$ in ~(\ref{tau3}) with
$N^{-1/3}$.

Following similar arguments for a scale-free network with degree distribution
$g(k)\sim k^{-\lambda}$ and $3<\lambda<4$, we can replace $L^{-1/\nu}$ by
$N^{-(\lambda-3)/(\lambda-1)}$ since
$d_c=2(\lambda-1)/(\lambda-3)$~\cite{cohen}. Thus,
\begin{equation}
S\sim AL^{-1/\nu}=AN^{-1/(d_c\nu)}=AN^{-(\lambda-3)/(\lambda-1)}.
\label{sf}
\end{equation}
For $\lambda>4$, Eq.~(\ref{sf}) reduces to ~(\ref{er})~\cite{cohen}.

Our main results, Eqs.~(\ref{tau3}), (\ref{er}) and (\ref{sf}), suggest that
since $S$ controls the disorder, for the same $S$, we expect to obtain the
same optimal path behavior independent of $P(w)$.

Next we calculate $A$ for several specific weight distributions $P(w)$. We
begin with the well-studied exponential disorder function $f(x)=e^{ax}$,
where $x$ is a random number between $0$ and $1$~\cite{strelniker,cieplak}.
From Eq.~(\ref{fx}) follows that $P(w)=1/(aw)$, where $w \in [1,e^a]$. Using
Eq.~(\ref{tau4}) we obtain $w_c=e^{ap_c}$, and from Eq.~(\ref{tau3}), $A$
becomes
\begin{equation}
A=ap_c
\label{exp}
\end{equation}
For fixed $S$, we expect to obtain the same optimal path behavior. Indeed,
this has been shown to be valid~\cite{zhenhua, perlsman, strelniker}.

Next we study $A$ for the disorder function $f(x)=x^{a}$, with $x$ between
$0$ and $1$, where $a$ is a parameter which can be positive or
negative~\cite{alex}. For this case the disorder distribution is a power law
$P(w)=\vert a\vert^{-1}w^{1/a-1}$. Following Eqs.~(\ref{tau3}) and
(\ref{tau4}), we obtain
\begin{equation}
A = |a|
\end{equation}
Note that $a$ plays a similar role to $a$ in Eq.~(\ref{exp}), but here $S$ is
independent of $p_c$, which means that networks with different $p_c$, such as
ER networks with different average degree $<k>=1/p_c$, will yield the same
optimal path behavior.

We further generalize the power law distribution with the disorder function
$f(x)=x^a$ by introducing the parameter $0\le\Delta\le 1$ which is defined as
the lower bound of the uniformly distributed random number $x$, i.e.,
$1-\Delta\le x\le 1$.  Under this condition, the distribution becomes
\begin{equation}
P(w)=\frac{w^{1/a-1}}{\vert a\vert \Delta}  \qquad   w\in [(1-\Delta)^a,1].
\end{equation}
Again using Eqs.~(\ref{tau3}) and (\ref{tau4}),we obtain
\begin{equation}
A= \frac{\Delta a p_c }{\Delta p_c+1-\Delta}.
\end{equation}

Table~\ref{table1} shows the results of similar analyses for the lognormal,
Gaussian, uniform and exponential distributions $P(w)$. From
Table~\ref{table1}, we see that for exponential function, power law and
lognormal distributions, $A$ and therefore $S$ can become large. However for
uniform, Gaussian and exponential distributions, $A$ is limited to a value of
order $1$, so $S\ll 1$ for large $L$. Note that for these distributions, $A$
is independent of $a$ or $\sigma$. Thus only the former distributions
(exponential function, power law, lognormal) are broad enough, and can lead
to strong disorder behavior for the optimal path, while for the latter
distributions (uniform, Gaussian and exponential), $S$ is always small and
only weak disorder can appear.

To test the validity of our theory, we perform simulations of optimal paths
in $2d$ square lattices and ER networks. Random weights from different
disorder functions were assigned to the bonds. For an $L\times L$ square
lattice, we calculate the average length $\ell$ of the optimal path from one
lattice edge to the opposite. For an ER network of $N$ nodes, we calculate
$\ell$ between two randomly selected nodes.

Figure 1 shows our simulation results for a $2d$ square lattice for power law
and lognormal distributions.  For small $L$, $\ell/L^{1.22}$ approachs a
constant, indicating the strong disorder regime $\ell \sim L^{1.22}$, while
for large $L$ it approaches to $L^{-0.22}$ indicating $\ell \sim L$. The
scaled results shown in Fig.1b show collapse for the same $S$ for all
distributions and therefore support Eq.~(\ref{tau3}) that $S$ controls the
optimal path, and that the optimal path $\ell$ behaves in the same universal
way for any broad distribution. Note that for large $S^{-\nu}$ the scaled
curves approach a slope of $1-d_{opt}\cong -0.22$ indicating the range of
weak disorder.

Scaled distributions $p(\ell)$ for general disorder distributions $P(w)$ are
shown in Fig. 2, in which $L$, $a$, $\Delta$ and $\sigma$ are selected to
keep $S$ constant. Figure 2 suggests that for the same value of $S$, the
distributions $p(\ell)$ collapse when plotted versus the scaling parameter
$\ell/A^{\nu d_{opt}}$.

Simulations for optimal paths on ER networks are shown in Fig. 3. Here we use
the bombing algorithm~\cite{cieplak,footnote} to determine the path length
$\ell_{\infty}$ in the strong disorder limit, which is related to $N$ by
$\ell_{\infty}\sim N^{\nu_{opt}}=N^{1/3}$~\cite{lidia,sameet}. We see that
for all disorder distributions studied, $\ell$ scales in the same universal
way with $S^{-1}\equiv \ell_{\infty}/A$. For $S \ll 1$, $\ell/A$ is linear
with $\mathrm{log}(\ell_{\infty}/A)$ as expected~(Fig. 3a). For large
$S=A/\ell _\infty$ (Fig. 3b), $\ell \propto \ell_{\infty}\sim N^{1/3}$, which
is the strong disorder behavior~\cite{lidia}. Thus, we see that when $N$
increases, a crossover from strong to weak disorder occurs in the scaled
optimal paths $\ell/A$ vs. $S^{-1}$. Again, the collapse of all curves for
different disorder distributions for ER networks supports the general
condition of Eq.~(\ref{tau3}).

Next we use Eq.~(\ref{tau3}) to analyze the other types of disorder given in
Table~\ref{table1} that do not have strong disorder behavior. For a uniform
distribution, $P(w)=1/a$ and we obtain $A=1$. The parameter $a$ cancels, so
$S=L^{-1/\nu}$ for lattices, and $S=N^{-1/3}$ for ER networks. Hence for any
value of $a$, $S\ll 1$, and strong disorder behavior cannot occur for a
uniform distribution.

Next we analyze the Gaussian distribution. We assume that all the weights
$w_j$ are positive and thus we consider only the positive regime of the
distribution. We obtain $A$ as given in Table I.  The disorder is controlled
solely by $p_c$ which is related only to the type of network, and $A$ cannot
obtain large values. Thus, also for the Gaussian $P(w)$, all optimal paths
are in the weak disorder regime.

Similar considerations lead to the same conclusion for the exponential
distribution.  Simulation results for the Gaussian distribution shown in Fig.
4 display only weak disorder (i.e. no weak-strong disorder crossover), thus
supporting the above conclusions.

In summary, we find a criterion for the disorder strength $S$ on the optimal
path in weighted networks for general distributions $P(w)$. We find an
analytical expression, Eq.~(\ref{tau3}), which fully characterizes the
behavior of the optimal path. We simulated several distributions and the
results support our analytical prediction. It is plausible that the criterion
of Eq.~(\ref{tau3}) is valid also for other properties in weighted networks
--- such as conductivity and flow in random resistor networks --- due to a
recently-found close relation between the optimal path and
flow~\cite{strelniker,zhenhua}.

Our results also suggest the conjecture that disorder distributions fall into
two classes: those $P(w)$ that can lead to strong disorder and those that
cannot. Our studies suggest that the ratio $\sigma/\mu$ determines the class,
where $\sigma^2$ is the variance of $P(w)$ and $\mu$ the mean value. If
$\sigma/\mu$ can become large ($\gg 1$), strong disorder can occur, but if
$\sigma/\mu\approx 1$ only weak disorder occurs. This criterion is valid for
all the cases simulated and we conjecture that it is valid for any weighted
distribution.

We thank ONR, European NEST project DYSONET, and Israel Science Foundation
for support.

%CONCLUSION
%}

%BIBLIOGRAPHY

\newpage

\begin{table}[h!]
\caption{Parameters controlling the optimal paths on networks for
  various distributions of disorder.}
\begin{center}
\begin{tabular}{|c|c|c|c|}
  \hline
  Name & Function & Distribution & $A$ \\
  \hline
  Inverse & $e^{ax}$ & $\frac{1}{aw} \ w \in [1,e^a)$ & $ap_c$\\
  \hline
  Power Law & $x^a$ & $\frac{w^{1/a-1}}{\vert a\vert\Delta}$  & $\frac{\Delta a p_c}{\Delta p_c +1-\Delta}$\\
  \hline
  Lognormal & $e^{\sqrt{2}\sigma \mathrm{erf^{-1}}(2x-1)}$  & $\frac{e^{-(\mathrm{ln}w)^2/2\sigma^2}}{w\sigma \sqrt{2\pi}}$
  & $\frac{\sqrt{2\pi} p_c\sigma}{e^{-[\mathrm{erf^{-1}}(2p_c-1)]^2}}$\\
  \hline
  Uniform & $ax$ & $1/a$ & $1$\\
  \hline
  Gaussian & $\sqrt{2}\sigma\mathrm{erf^{-1}}(x)$ &
  $\frac{2e^{-w^2/(2\sigma^2)}}{\sigma\sqrt{2\pi}}$
  & $\frac{\sqrt{\pi}p_c e^{[\mathrm{erf^{-1}}(p_c)]^2}
  }{2\mathrm{erf^{-1}}(p_c)}$\\
  \hline
  Exponential & $-\frac{\mathrm{ln}(1-x)}{\zeta}$  & $\zeta e^{-\zeta w}$ &
  $\frac{p_c}{(p_c-1)\mathrm{ln}(1-p_c)}$\\
  \hline
\end{tabular}
\end{center}
\label{table1}
\end{table}

\newpage

\begin{figure}[h!]
  \centerline{
    \epsfysize=0.43\columnwidth{\rotatebox{0}{\epsfbox{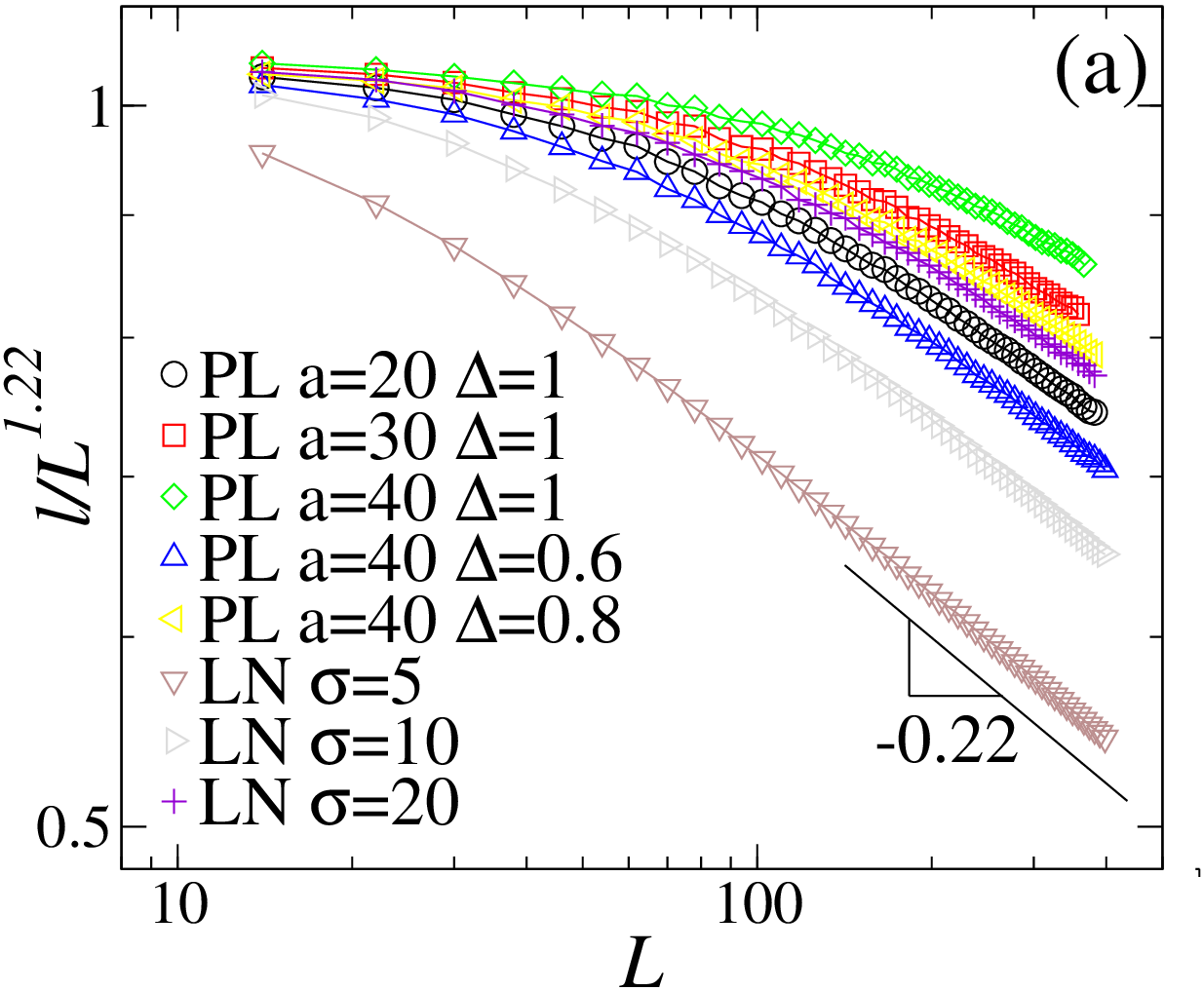}}}
    \epsfysize=0.43\columnwidth{\rotatebox{0}{\epsfbox{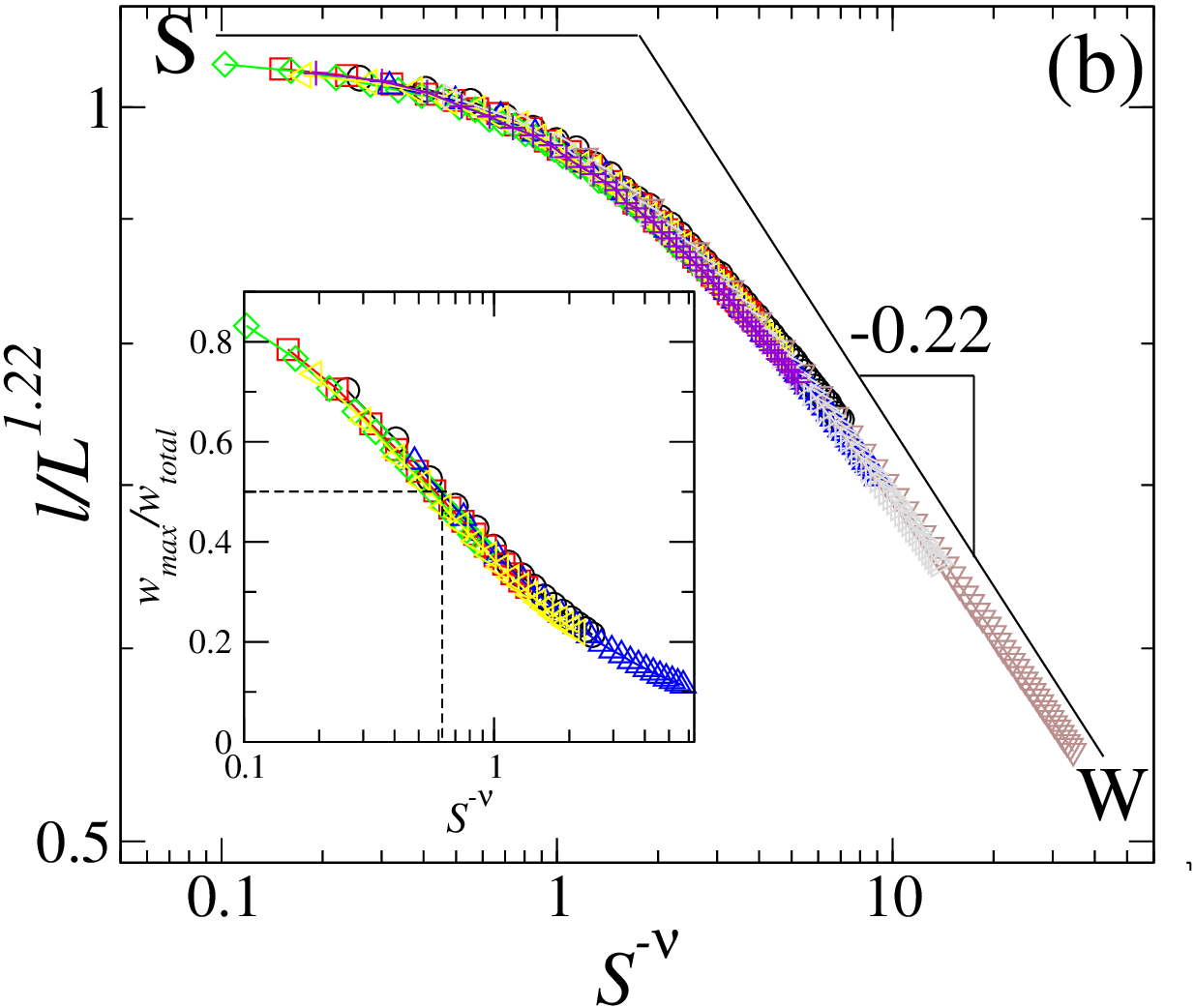}}}
  }
  \caption{Simulation results for different disorder distributions in $2d$
    lattice $(a)$ before and $(b)$ after scaling. A lower limit of $L=15$ is
    set to avoid the finite size effects. $(a)$ Curves for power law disorder
    chosen from $x^a$, $0\le x<1$, with $a$ from $20$ to $40$, fixing
    $a=40$ and $\Delta=0.6$ or $0.8$, and lognormal distribution with
    $\sigma$ from $5$ to $20$. $(b)$ Same results plotted in a scaled form.
    Two lines are drawn in $(b)$, with slopes $0$ and $-0.22$, indicating the
    theoretical behavior of optimal path in the strong and weak disorder
    limits respectively. The inset in (b) shows the ratio between the highest
    weight and total weight versus $S^{-\nu}$. A line of
    $w_{max}/w_{total}=0.5$ indicates the point $S^*$ above which the highest
    weight dominates the total weight.}
\label{graph1}
\end{figure}

\newpage

\begin{figure}[h!]
  \centerline{
    \epsfysize=0.43\columnwidth{\rotatebox{0}{\epsfbox{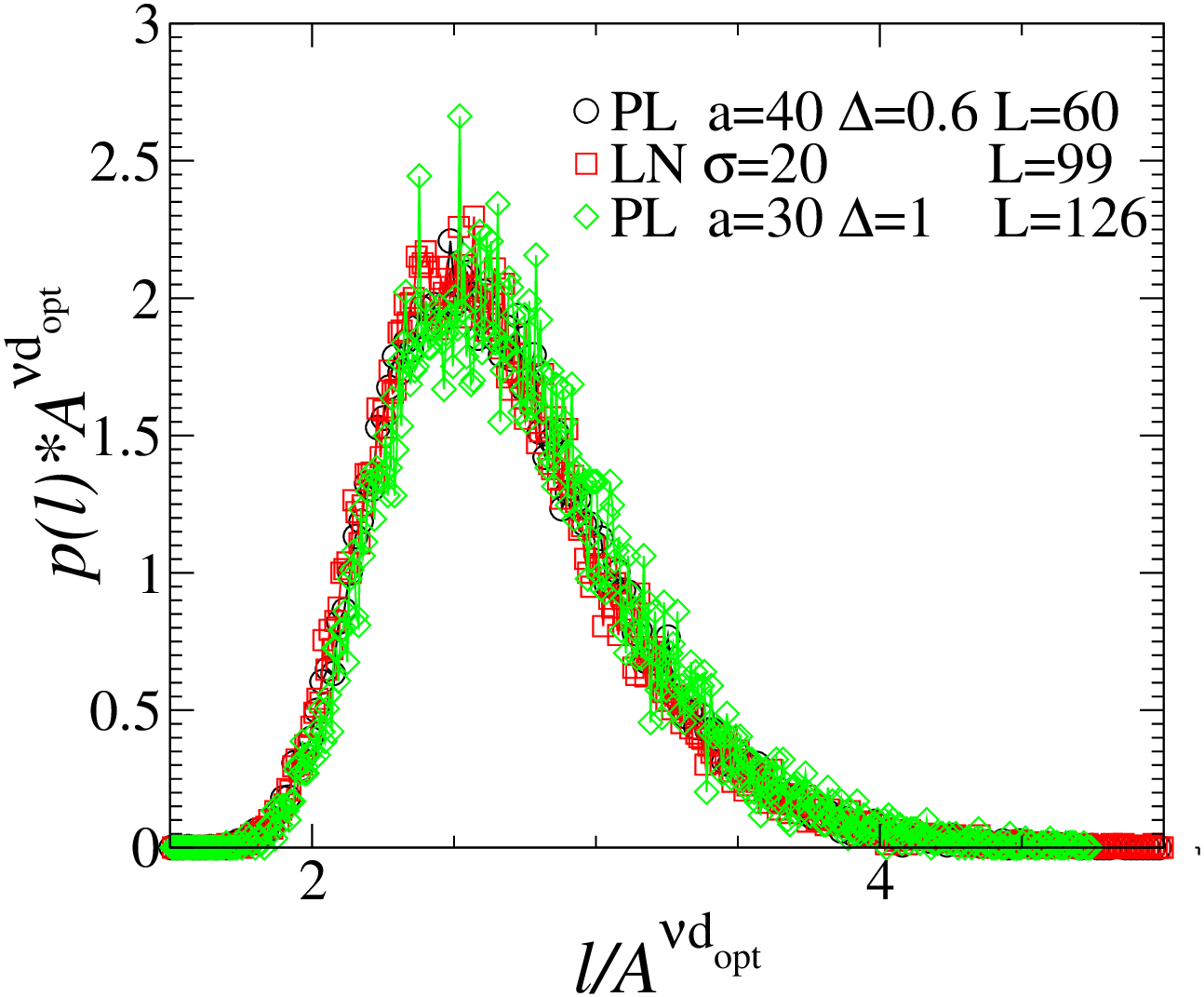}}}
}

\caption{Scaled distributions of $p(\ell)$ in $2d$
  lattice. Distributions used here are lognormal distribution with
  $\sigma=20$, power law distribution with $a=30$ and power law distribution
  with $a=40, \Delta=0.6$. The linear size $L$ of lattice is selected by
  fixing $S=0.8$.}
\label{graph2}
\end{figure}

\newpage

\begin{figure}[h!]
  \centerline{
    \epsfysize=0.43\columnwidth{\rotatebox{0}{\epsfbox{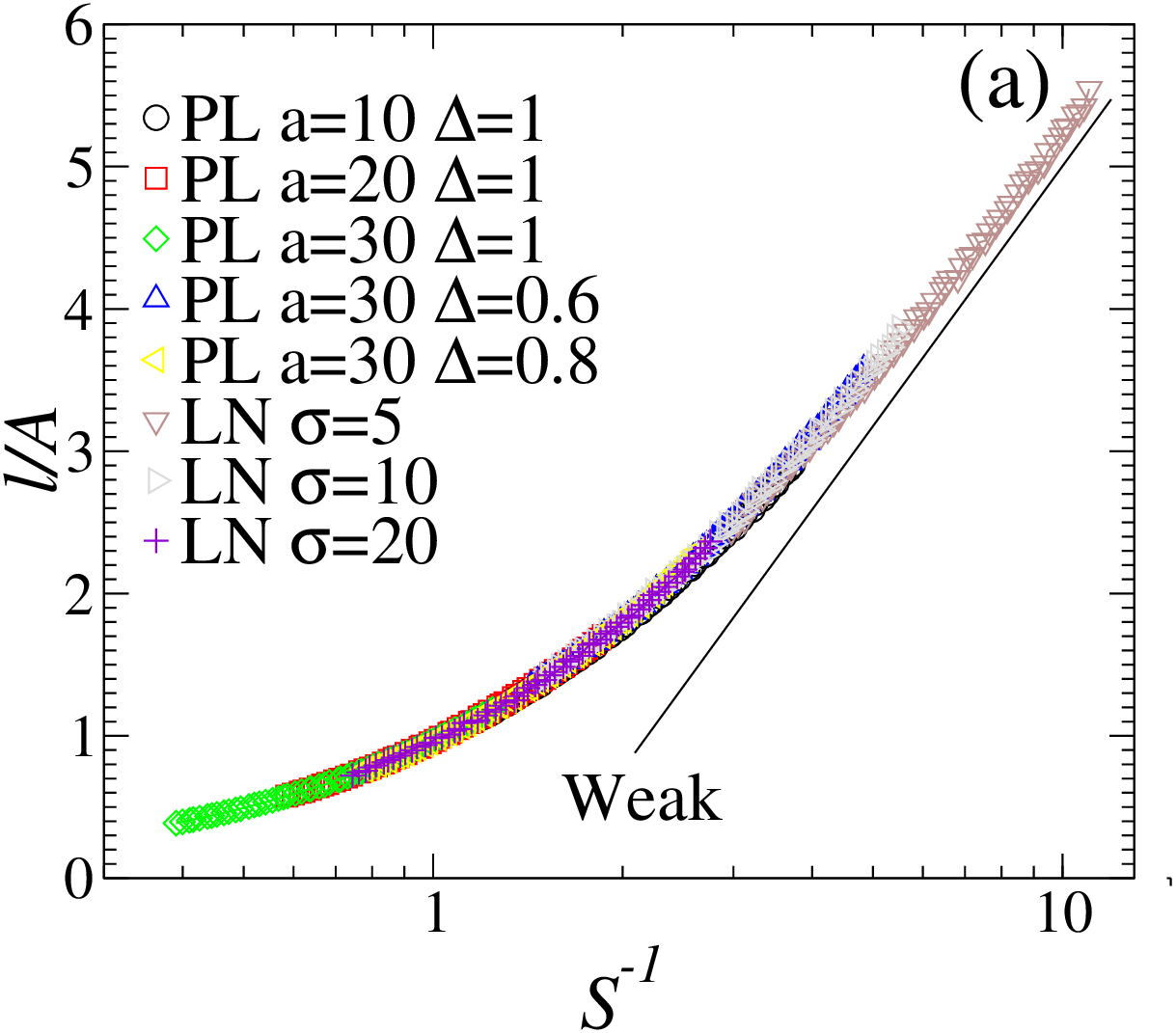}}}
    \epsfysize=0.43\columnwidth{\rotatebox{0}{\epsfbox{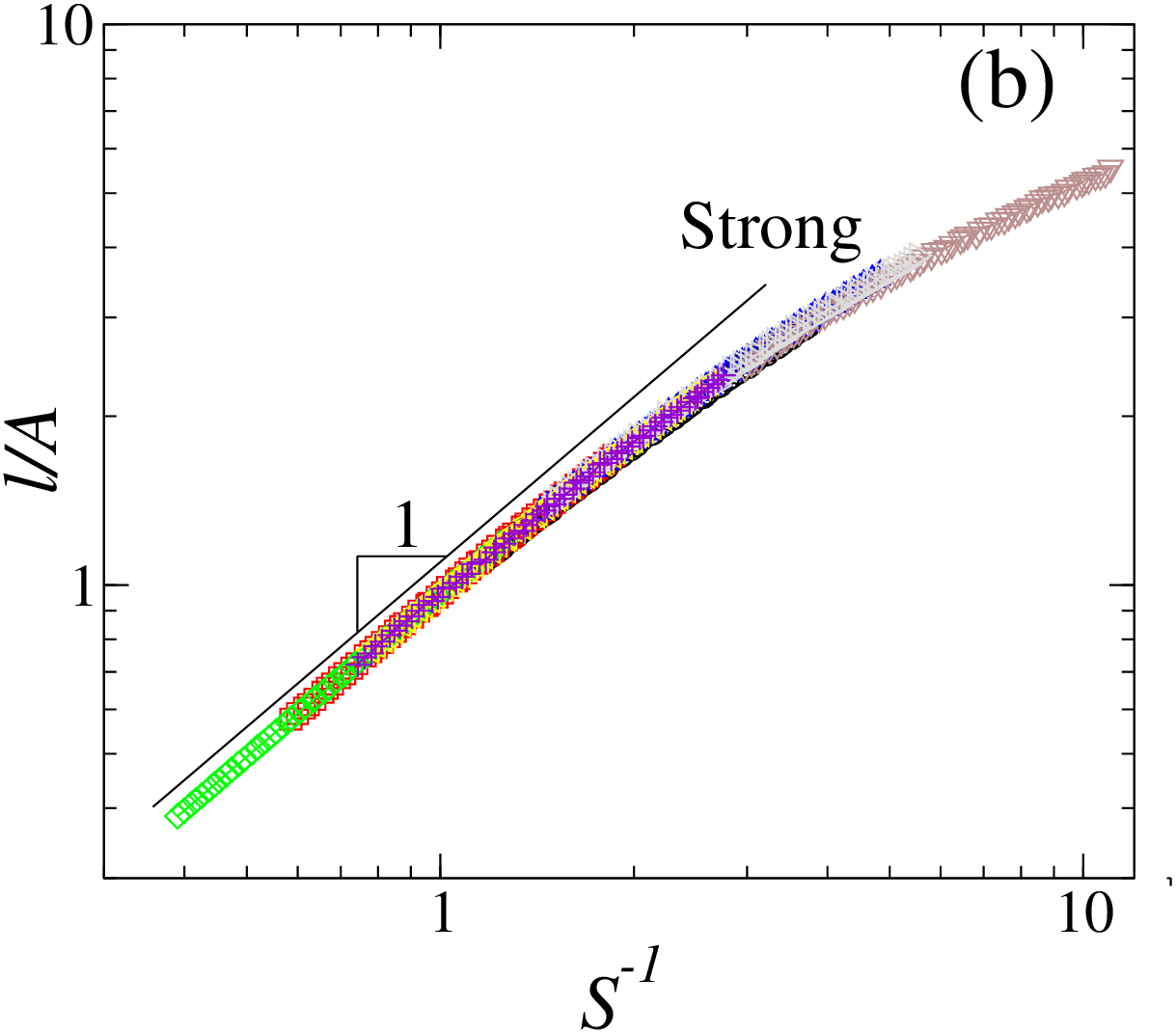}}}
}
\caption{The function $\ell/A$ for ER networks after scaling, where $(a)$ is
  a linear-log plot and $(b)$ is a log-log plot. Distributions used are power
  law $x^a$ with $10\le a\le 30$ where $0\le x<1$, $x^a$ with $a=30$ and the
  range of $\Delta<x\le 1$ with $\Delta=0.6$ or $0.8$, and lognormal
  distribution with $10\le \sigma \le 30$. The straight line in $(a)$
  indicates weak disorder and the straight line in $(b)$ indicates strong
  disorder. }
\label{graph3}
\end{figure}
%\begin{figure}
%\centerline{
%  \epsfysize=0.4\columnwidth{\rotatebox{-90}{\epsfbox{gaussian2}}}
%}
%\label{gaussian}
%\caption{The relation between $S$ and $T$ when $p_c=0.5$. We can see
%  $S$ is asymptotically go to a constant when $T$ goes negative infinite
%  and $0$ when $T$ goes positive infinite}
%\end{figure}

\newpage

\begin{figure}[h!]
\centerline{
  \epsfysize=0.43\columnwidth{\rotatebox{0}{\epsfbox{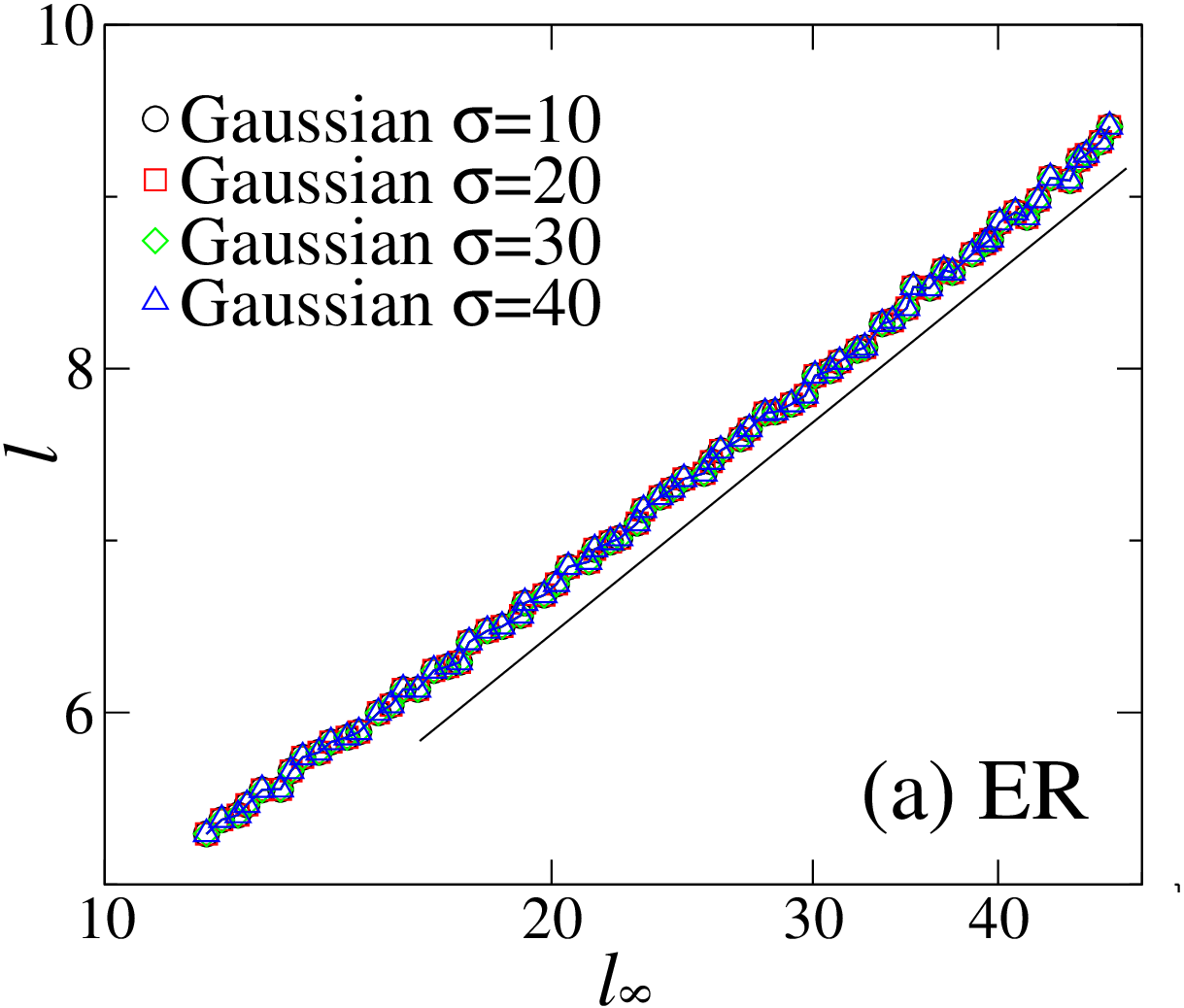}}}
  \epsfysize=0.43\columnwidth{\rotatebox{0}{\epsfbox{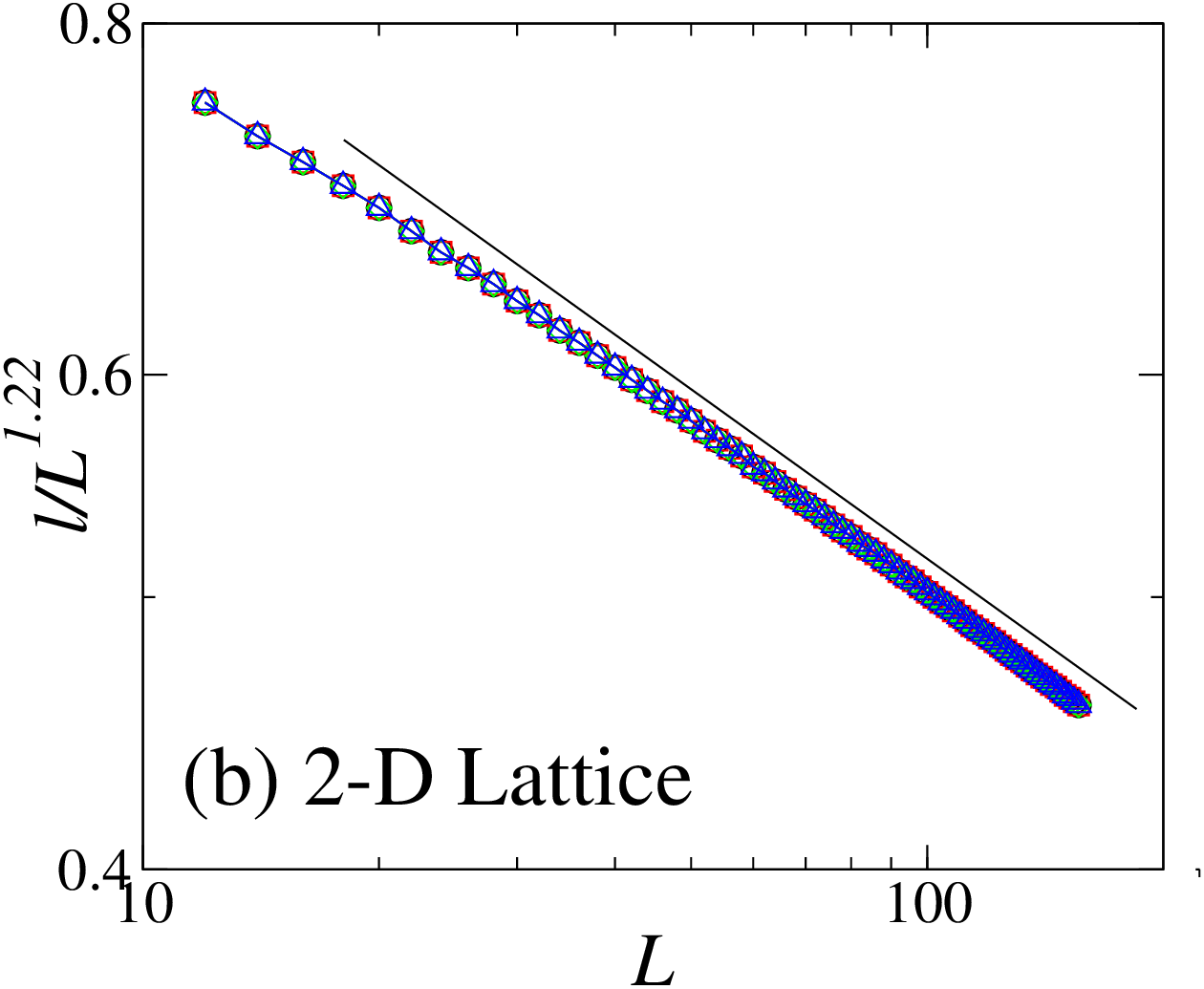}}}
}
\caption{The optimal path for Gaussian distribution of weights. $(a)$
  Gaussian distribution for ER networks and $(b)$ Gaussian distribution for
  $2d$ lattices. Note that these curves would collapse after scaling to the
  curves in figure 1b in the weak disorder tail of large $S^{-\nu}$.}
\label{graph4}
\end{figure}

\begin{thebibliography}{99}
\bibitem{rintoul} M.D. Rintoul, J. Moon, and H. Nakanishi, Phys. Rev. E
  \textbf{49}, 2790 (1994).
\bibitem{cieplak} M. Cieplak, A. Maritan, and J.R. Banavar,
  Phys. Rev. Lett. \textbf{72}, 2320 (1994).
\bibitem{strelniker} Y. M. Strelniker et al., Phys. Rev. E \textbf{69},
  065105(R) (2004).
\bibitem{barabasi-stanley} A. -L. Barab\'{a}si and H. E. Stanley,
{\it Fractal Concepts in Surface Growth} (Cambridge, NY, 1995).
\bibitem{void}M. Mezard, G. Parisi, M.A. Virasoo, {\it Spin Glass Theory and
    Beyond} (World Scientific Pub, 1987).
\bibitem{bru} A. Br\'{u}, S. Albertos, J.A.L\'{o}pez Garc\'{i}a-Asenjo, and
  I. Br\'{u}, Phys. Rev. Lett. \textbf{92}, 238101 (2004).
\bibitem{lidia} L.A. Braunstein et al.  Phys. Rev. Lett. \textbf{91}, 168701
  (2003).
\bibitem{alex} A. Hansen and J. Kert\'{e}sz, Phys. Rev. Lett. \textbf{93},
  040601 (2004).
\bibitem{edward} E. McCann and I.V. Lerner, J. Phys. Cond. Matt. \textbf{8},
  6719 (1996).
\bibitem{redner} I. Smailer et al., Phys. Rev. E
  \textbf{47}, 262 (1993)
\bibitem{footnote1} This criterion can be regarded as the definition of strong
  disorder, see also [2].
\bibitem{porto} M. Porto et al., Phys. Rev. E
  \textbf{60}, R2448 (1999).
\bibitem{footnote} This is actually the bombing algorithm that determines the
  optimal path in strong disorder~\cite{cieplak}.
\bibitem{stauffer} D. Stauffer and A. Aharony, {\it Introduction to
    Percolation Theory} (Taylor \& Francis, London, 1994).
\bibitem{bunde} A. Bunde and S. Havlin, {\it Fractals and Disordered System}
  (Springer, Berlin, 1991).
\bibitem{erdos} P. Erd\H{o}s and A. R\'{e}nyi, Publ. Math. (Debrecen)
  \textbf{6}, 290 (1959).
\bibitem{cohen} R. Cohen et al., Phys. Rev. E \textbf{66}, 036113 (2002).
\bibitem{perlsman} E. Perlsman and S. Havlin, Eur. Phys. J. B \textbf{43}
  517 (2005).
\bibitem{zhenhua} Z. Wu et al., Phys. Rev. E \textbf{71}, 045101(R) (2005).
\bibitem{sameet} S. Sreenivasan et al., Phys. Rev. E \textbf{70}, 046133
  (2004).


\end{thebibliography}
\end{document}